\documentclass[11pt,fleqn]{article} 
\usepackage{psfig}
\title{Possible Routes to  Frictionless Transport of Electronic \\ Fluids in High-Temperature Superconductors}  %
\author{Zotin K.-H. Chu} 
\date{3/F, 24, 260th. Lane, First Section, Muja Road, Taipei, Taiwan 116,
China and\\ P.O. Box 39, Distribution Unit, Xihong Road, Urumqi
830000, China}
\begin{document}           
\maketitle
\doublerulesep=6.5mm        
\baselineskip=6.5mm
\oddsidemargin-1mm         
\begin{abstract}
Electric-field-driven transport of electronic fluids in metallic
glasses as well as three-dimensional amorphous superconductors are
investigated by using the verified approach which has been
successfully adopted  to study the critical transport of glassy
solid helium in very low temperature environment. The critical
temperatures related to the nearly frictionless transport of
electronic fluids were found to be directly relevant to the
superconducting temperature of amorphous superconductors after
selecting specific activation energies. Our results imply that
optimal shear-thinning is an effective way to reach
high-temperature charged superfluidity or
superconductivity.\newline

\end{abstract}
%
\bibliographystyle{plain}
\section{Introduction}
Analogies between superconductivity (charged superfluidity) and
superfluidity in liquid [1-2] and solid helium (either supersolid
[3] or superglass [4-6])  has been noted by some researchers. The
particles (charged electrons as well as atoms) in a superfluid are
in the same quantum state, so they move coherently and result in
dissipationless mass flow. Quite recently a glassy state of
supersolid helium (solid $^4$He : ability to flow without
resistance) has been reported [5]. We have made some contributions
to the possible transport of  glassy solid $^4$He in confined
domain under rather low temperature environment (cf. Ref. 6 for
the comparison with available relevant experiments). Our present
interest is to borrow the above mentioned analogy to investigate
the nearly frictionless states for transport of the electronic
liquid (presumed to be glassy or amorphous) in amorphous
superconductors [7-10] which is closely related to charged
frictionless or superfluidity states. The approach adopted in [6]
(considering transport of glassy fluids) will be extended by
varying the forcing to be an electric field in this paper.
\newline
Note  that one prominent difference between the fluid transport in
microdomain and those in macrodomain is the strong fluid-wall
interactions observed in microconduits. For example, as the
microconduit size decreases, the surface-to-volume ratio
increases. Therefore, various properties of the walls, such as
surface roughness, greatly affect the fluid motions in
microconduits.
%
\newline
In this  paper,  we adopt the verified Eyring model [6,11-13] to
study the nearly frictionless transport of (glassy) electronic
fluids within corrugated microannuli. Eyring model treats
transport of shear-thinning (which means once material is
subjected to high shear rates then the viscosity diminishes with
increasing shear rate) matter quite successfully [6,11-13]. To
obtain the law of shear-thinning fluids for explaining the too
rapid annealing at the earliest time, because the relaxation at
the beginning was steeper than could be explained by the
bimolecular law, a hyperbolic sine law between the shear (strain)
rate : $\dot{\gamma}$ and (large) shear stress : $\tau$ was
proposed and the close agreement with experimental data was
obtained [11-13]. This model has sound physical foundation from
the thermal activation process  (Eyring [12] already considered a
kind of (quantum) tunneling which relates to the matter
rearranging by surmounting a potential energy barrier). With this
model we can associate the (shear-thinning) electronic fluid with
the momentum transfer between neighboring  clusters on the
microscopic scale and reveals the detailed interaction in the
relaxation of flow with dissipation (the momentum transfer depends
on the activation shear volume, which is associated with the
center distance between (charged) particles and is proportional to
$k_B T/\tau_0$ ($T$ is temperature in Kelvin, and $\tau_0$ a
constant with the dimension of stress). Thus, this model could be
applied to study transport of (shear-thinning as well as amorphous
or glassy) electronic fluids in microdomain which is of particular
interest to researchers in condensed matter physics (considering
amorphous superconductors [7-10]). We select the microannular
geometry as it is more relevant to the experimental environment
considering the measuring conductivity of samples of amorphous
superconductors [7-10] (transport of shell-like electronic
fluids).
\newline To consider the more realistic but complicated  boundary
conditions in the walls of microannulus, however, we will use the
boundary perturbation technique [13-14] to handle the presumed
wavy-roughness along the walls of microannuli. To obtain the
analytical and approximate solutions, here, the roughness is only
introduced in the radial or transverse direction. The relevant
boundary conditions along the wavy-rough surfaces will be
prescribed below. We shall describe our approach after this
section : Introduction with the focus upon the boundary
perturbation method. The expression of the electric-field-driven
volume flow rate of the glassy (or amorphous) electronic fluid is
then demonstrated at the end. We noticed that the conductivity
measurements reported in [9] and [10] are primary in Mg-based and
Fe$_x$Ni$_{1-x}$-based metallic glasses. Finally, we will
illustrate our results into four figures and give discussions
therein. Our results show that some specific activation energies
and activation volumes are crucial to the nearly frictionless
states of the transport of the electronic fluids at rather low
temperature regime. The good qualitative as well as quantitative
(e.g., critical temperature or $T_c$) comparison with experiments
confirms our approaches. Our results imply that optimal
shear-thinning is an effective way to reach high-temperature
charged superfluidity or superconductivity.
\section{Formulations} %
Researchers have been interested in the question of how amorphous
material responds to an external mechanical load. External loads
cause (electronic) liquids to flow, in Newtonian or various types
of non-Newtonian flows. Glassy materials, composed of polymers,
metals, or ceramics, can deform under mechanical loads, and the
nature of the response to loads often dictates the choice of
material in various industrial applications.  The nature of all of
these responses depends on both the temperature and loading rate.
As described by Eyring \cite{Eyr:JCP-4}, mechanical loading lowers
energy barriers, thus facilitating progress over the barrier by
random thermal fluctuations. The Eyring model approximates the
loading dependence of the barrier height as linear. The Eyring
model, with this linear barrier height dependence on load, has
been used over a large fraction of the last century to describe
the response of a wide range of systems  and underlies modern
approaches to sheared glasses [11-12]. To the best knowledge of
the author, the simplest model that makes a prediction for the
rate and temperature dependence of shear yielding is the
rate-state Eyring model of stress-biased thermal activation
[11-12]. Structural rearrangement is associated with a single
energy barrier $E$ that is lowered or raised linearly by an
applied stress $\sigma$.
$$R_{\pm}=\nu_0 \exp[-E/(k_B T)] \exp[\pm \sigma V^*/(k_B T)],$$
where $k_B$ is the Boltzmann constant, $\nu_0$ is an attempt
frequency and $V^*$ is a constant called the activation volume. In
glasses, the transition rates are negligible at zero stress. Thus,
at finite stress one needs to consider only the rate $R_{+}$ of
transitions in the direction aided by stress.\newline The linear
dependence will always correctly describe small changes in the
barrier height, since it is simply the first term in the Taylor
expansion of the barrier height as a function of load. It is thus
appropriate when the barrier height changes only slightly before
the system escapes the local energy minimum. This situation occurs
at higher temperatures; for example, Newtonian flow is obtained in
the Eyring model in the limit where the system experiences only
small changes in the barrier height before thermally escaping the
energy minimum. As the temperature decreases, larger changes in
the barrier height occur before the system escapes the energy
minimum (giving rise to, for example, non-Newtonian flow). In this
regime, the linear dependence is not necessarily appropriate, and
can lead to inaccurate modeling. To be precise, at low shear rates
($\dot{\gamma} \le \dot{\gamma}_c$), the system behaves as a power
law shear-thinning material while, at high shear rates, the stress
varies affinely with the shear rate. These two regimes correspond
to two stable branches of stationary states, for which data
obtained by imposing either $\sigma$ or $\dot{\gamma}$ exactly
superpose. The transition from the lower branch to the higher
branch occurs through a stable hysteretic loop in a
stress-controlled experiment.
\newline
The motion (of composite particles) is represented in the
configuration space; on the potential surface the stable particles
are in the valleys, which are connected by a pass that leads
through the saddle point. A (composite) particle at the saddle
point is in the transition (activated) state. Under the action of
an applied stress the forward velocity of a  flow unit is the net
number of times it moves forward, multiplied by the distance it
jumps. Eyring proposed a specific microscopic model of the
amorphous structure and a mechanism of deformation kinetics
[11-12]. With reference to this idea, this mechanism results in a
(shear) strain rate given by
\begin{equation}  
 \dot{\gamma}=2\frac{V_h}{V_m}\frac{k_B T}{h}\exp (\frac{-\Delta
E}{k_B T}) \sinh(\frac{V_h \tau}{2 k_B T})
\end{equation}
where $$V_h=\lambda_2\lambda_3\lambda, \hspace*{24mm}
V_m=\lambda_2\lambda_3\lambda_1,$$ $\lambda_1$ is the
perpendicular distance between two neighboring layers of
(composite) particles sliding past each other, $\lambda$ is the
average distance between equilibrium positions in the direction of
motion, $\lambda_2$ is the distance between neighboring
(composite) particles in this same direction (which may or may not
equal $\lambda$), $\lambda_3$ is the (composite) particle to
(composite) particle distance in the plane normal to the direction
of motion, and $\tau$ is the local applied stress, $\Delta E$ is
the activation energy, $h$ is the Planck constant, $k_B$ is the
Boltzmann constant, $T$ is the temperature, $V_h$ is the
activation volume for the microscopic event [12]. The deformation
kinetics of the chain (composite particles) is envisaged as the
propagation of kinks in the particles into available holes. In
order for the motion of the kink to result in a  flow, it must be
raised (energised) into the activated state and pass over the
saddle point. This was the earliest microscopic theory of yield
behaviour in amorphous materials, and Eyring presented a
theoretical framework which formed the basis of many subsequent
considerations.
\newline
Solving Eqn. (1) for  $\tau$, one obtains:
\begin{equation}
  \tau=\frac{2 k_B T}{V_h} \sinh^{-1} (\frac{\dot{\gamma}}{B}),
\end{equation}
which in the limit of small $(\dot{\gamma}/B)$ reduces to Newton's
law for viscous deformation kinetics.
\newline
We shall consider a steady transport of the glassy (or amorphous)
electronic fluids in a wavy-rough microannulus of $r_1$
(mean-averaged inner radius) with the inner interface being a
fixed wavy-rough surface : $r=r_1+\epsilon \sin(k \theta+\beta)$
and $r_2$ (mean-averaged outer radius) with the outer interface
being a fixed wavy-rough surface : $r=r_2+\epsilon \sin(k
\theta)$, where $\epsilon$ is the amplitude of the (wavy)
roughness, $\beta$ is the phase shift between two boundaries, and
the roughness wave number : $k=2\pi /L $ ($L$ is the wavelength of
the surface modulation in
transverse direction). 
\newline Firstly, this amorphous fluid [6,12-13] can be expressed as
 $\dot{\gamma}=\dot{\gamma}_0  \sinh(\tau/\tau_0)$,
where $\dot{\gamma}$ is the shear rate, $\tau$ is the shear
stress, and $\dot{\gamma}_0 \equiv B$ is a function of temperature
with the dimension of the shear rate. In fact, the force balance
gives the shear stress at a radius $r$ as $\tau=-[r \,\delta
(\rho_e E_z)]/2$, where $\tau$ is the shear stress along the
boundaries of a control volume in the flow direction (the same
direction as $E_z$ which (the only electric  field) is presumed to
be a constant or uniform), $\rho_e$ is the net charge density,
$|-\delta(\rho_e E_z)|$ is the net electric force along the
transport (or tube-axis : $z$-axis) direction.\newline Introducing
$\chi = -(r_2/2\tau_0) \delta(\rho_e E_z)$
then we have
 $\dot{\gamma}= \dot{\gamma}_0  \sinh ({\chi r}/{r_2})$.
As $\dot{\gamma}=- du/dr$ ($u$ is the velocity of the (electronic)
fluid transport in the longitudinal ($z$-)direction of the
microannulus), after integration, we obtain
\begin{equation}
 u=u_s +\frac{\dot{\gamma}_0 r_2}{\chi} [\cosh \chi - \cosh (\frac{\chi r}{r_2})],
\end{equation}
here, $u_s$ is the velocity over the (inner or outer) surface of
the microannulus, which is determined by the boundary condition.
We noticed that  a general boundary condition [13] was proposed
for transport over an interface as
\begin{equation}
 \Delta u=L_s^0 \dot{\gamma}
 (1-\frac{\dot{\gamma}}{\dot{\gamma}_c})^{-1/2},
\end{equation}
where  $\Delta u$ is the velocity jump over the interface, $L_s^0$
is a constant slip length, $\dot{\gamma}_c$ is the critical shear
rate at which the slip length diverges. The value of
$\dot{\gamma}_c$ is a function of the corrugation of interfacial
energy.  \newline With the boundary condition from  [13],  we can
derive the velocity fields and electric-field-driven volume flow
rates along the wavy-rough microannulus below using the verified
boundary perturbation technique [6,13-14]. The wavy boundaries are
prescribed as $r=r_2+\epsilon \sin(k\theta)$ and $r=r_1+\epsilon
\sin(k\theta+\beta)$ and the presumed steady transport is along
the $z$-direction (microannulus-axis direction).
\newline Along the outer boundary (the same treatment below could also be
applied to
the inner boundary), we have
 $\dot{\gamma}=(d u)/(d n)|_{{\mbox{\small on surface}}}$.
Here, $n$ means the  normal. Let $u$ be expanded in $\epsilon$ :
 $u= u_0 +\epsilon u_1 + \epsilon^2 u_2 + \cdots$,
and on the boundary, we expand $u(r_0+\epsilon dr,
\theta(=\theta_0))$ into
\begin{displaymath}
u(r,\theta) |_{(r_0+\epsilon dr,\,\theta_0)} =u(r_0,\theta)+\epsilon
[dr \,u_r (r_0,\theta)]+ \epsilon^2 [\frac{dr^2}{2}
u_{rr}(r_0,\theta)]+\cdots=
\end{displaymath}
\begin{equation}
  \{u_{slip} +\frac{\dot{\gamma} r_2}{\chi} [\cosh \chi - \cosh (\frac{\chi
 r}{r_2})]\}|_{{\mbox{\small on surface}}}, \hspace*{6mm} r_0 \equiv
 r_1, r_2;
\end{equation}
where
\begin{equation}
 u_{slip}|_{{\mbox{\small on surface}}}=L_S^0 \{\dot{\gamma}
 [(1-\frac{\dot{\gamma}}{\dot{\gamma}_c})^{-1/2}]\}
 |_{{\mbox{\small on surface}}}. 
\end{equation}
Now, on the outer interface (cf. [13-14])
\begin{equation}   
 \dot{\gamma}=\frac{du}{dn}=\nabla u \cdot\frac{\nabla (r-r_2-\epsilon
\sin(k\theta))}{| \nabla (r-r_2-\epsilon \sin(k\theta))
|}.
\end{equation}
Considering $L_s^0 \sim r_1,r_2 \gg \epsilon$ case, we also presume
$\sinh\chi \ll \dot{\gamma}_c/\dot{\gamma_0}$.
With equations (1) and (5), using the definition of $\dot{\gamma}$,
we can derive the velocity field ($u$) up to the second order :
$u(r,\theta)$$=-(r_2 \dot{\gamma}_0/\chi) \{\cosh (\chi
r/r_2)$$-\cosh\chi\, [1+\epsilon^2 \chi^2 \sin^2 (k\theta)/(2
r_2^2)]+$$\epsilon \chi \sinh \chi \,
\sin(k\theta)/r_2\}$$+u_{slip}|_{r=r_2+\epsilon \sin
(k\theta)}$. The key point is to firstly obtain the slip
velocity along the boundaries or surfaces.
After lengthy mathematical manipulations, we obtain %
the velocity fields (up to the second order) and then we can
integrate them with respect to the cross-section to get the volume
flow rate ($Q$, also up to the second order here) :
%
 $ Q=\int_0^{\theta_p} \int_{r_1+\epsilon \sin(k\theta+\beta)}^{r_2+\epsilon \sin(k\theta)}
 u(r,\theta) r
 dr d\theta =Q_{slip} +\epsilon\,Q_{p_1}+\epsilon^2\,Q_{p_2}$.
In fact, the approximately (up to the second order) net
electric-field-driven volume flow rate  reads $Q \equiv Q_{out} -
Q_{in}$ which is the flow within the outer (larger) wall :
$Q_{out}$ without the contributions from the flow within the inner
(smaller) wall $Q_{in}$ :
\begin{displaymath}
 Q=\pi \dot{\gamma}_0 \{L_s^0 (r_2^2-r_1^2)  \sinh\chi \,
 (1-\frac{\sinh\chi}{\dot{\gamma}_c/\dot{\gamma_0}})^{-1/2}+
 \frac{r_2}{\chi}[(r_2^2-r_1^2)\cosh\chi-\frac{2}{\chi}(r_2^2 \sinh \chi-
\end{displaymath}
\begin{displaymath}
  r_1 r_2 \sinh(\chi \frac{r_1}{r_2})+ \frac{2r_2^2}{\chi^2}(\cosh\chi-\cosh(\chi \frac{r_1}{r_2})]\}+
  \epsilon^2 \{\pi\dot{\gamma}_0 [\chi \frac{\cosh \chi}{4}
  (r_2-\frac{r_1^2}{r_2})]+
\end{displaymath}
\begin{displaymath}
 \frac{\pi}{4} \dot{\gamma}_0  \sinh \chi (1+\frac{\sinh\chi}{\dot{\gamma}_c/\dot{\gamma_0}})
 (-k^2+\chi^2)[1-(\frac{r_1}{r_2})^2]+
 \frac{\pi}{2}  [(u_{slip_0} +
\end{displaymath}
\begin{displaymath}
\frac{\dot{\gamma}_0\,r_2}{\chi} \cosh\chi) +\dot{\gamma}_0 r_2(
  -\sinh\chi +\frac{\cosh\chi}{\chi})+ \dot{\gamma}_0(r_1 \sinh
  (\chi \frac{r_1}{r_2})-\frac{r_2}{\chi} \cosh (\chi
  \frac{r_1}{r_2}))]+
\end{displaymath}
\begin{displaymath}
  \pi \dot{\gamma}_0 \{[\sinh\chi+\chi\frac{\cosh\chi}{r_2}
  (1+\frac{\sinh\chi}{\dot{\gamma}_c/\dot{\gamma_0}})] [r_2-r_1 \cos\beta
  ]\}+
\end{displaymath}
\begin{equation}
  \frac{\pi}{4} \chi^2 \dot{\gamma}_0 \frac{\cosh\chi}{\dot{\gamma}_c/\dot{\gamma}_0}[1
-(\frac{r_1}{r_2})^2
 ]
 \cosh\chi.
\end{equation}
Here,
\begin{equation}
 u_{{slip}_0}= L_s^0 \dot{\gamma}_0 [\sinh\chi(1-\frac{\sinh\chi}{
 \dot{\gamma}_c/\dot{\gamma}_0})^{-1/2}].
\end{equation}
\section{Results and Discussions}
We firstly check the roughness effect upon the shearing
characteristics because there are no available experimental data
and numerical simulations for the same geometric configuration
(microscopic annuli with wavy corrugations in transverse
direction). With a series of forcings (due to imposed electric
fields) : $\chi\equiv r_2 [-\delta (\rho_e E_z)]/(2\tau_0)$, we
can determine the enhanced shear rates ($d\gamma/dt$) due to
forcings. From equation (7), we have (up to the first order)
\begin{equation}
 \frac{d\gamma}{dt}=\frac{d\gamma_0}{dt} [ \sinh \chi+\epsilon
 \sin(k\theta) \frac{\chi}{r_2} \cosh \chi].
\end{equation}
Furthermore, if we select a (fixed) temperature, then from the
expression of $\tau_0$, we can obtain the shear stress $\tau$
corresponding to above forcings ($\chi$) :
\begin{equation}
 \tau =\tau_0 \sinh^{-1} [\sinh(\chi)+\epsilon
 \sin(k\theta) \frac{\chi}{r_2} \cosh(\chi)].
\end{equation}
Note that, based on the absolute-reaction-rate Eyring model (of
stress-biased thermal activation), structural rearrangement is
associated with a single energy barrier (height) $\Delta E$ that
is lowered or raised linearly by a (shear) yield stress $\tau$. If
the transition rate is proportional to the plastic (shear) strain
rate (with a constant ratio : $C_0$; $\dot{\gamma} = C_0 R_t$,
$R_t$ is the transition rate in the direction aided by stress), we
have
\begin{equation}
 \tau = 2 [\frac{\Delta E}{V_h} + \frac{k_BT}{V_h} \ln(\frac{\dot{\gamma}}{C_0
 \nu_0})]  \hspace*{28mm} \mbox{if}\hspace*{6mm} \frac{V_h
 \tau}{k_B T} \gg 1
\end{equation}
where $\nu_0$ is an attempt frequency or transition rate, $C_0
\nu_0 \sim  \dot{\gamma}_0 \exp(\Delta E/k_B T)$, or
\begin{equation}
 \tau = 2\frac{k_BT}{V_h} \frac{\dot{\gamma}}{C_0
 \nu_0}\exp({\Delta E}/{k_B T})   \hspace*{24mm} \mbox{if}\hspace*{6mm} \frac{V_h
 \tau}{k_B T} \ll 1.
\end{equation}
The nonlinear character only manifests itself when the magnitude
of the applied stress times the activation volume becomes
comparable or greater in magnitude than the thermal vibrational
energy.
\newline
The following figures are for our primary interest : Higher
(electric) conductivity or lower viscosity which means rather low
resistance of the electric-field-driven transport of amorphous
 electron liquid.
This can be manifested from equation (8) for the net transport
$Q\, (\propto \sinh \chi)$. We have the average velocity (for
charge carriers) : $\bar{v}=Q/A_m$ with $A_m$  being the effective
area. After this, we then have the electric flux : $\rho_e
\bar{v}=|{\bf J}|$ which conventionally equals to $\sigma E_z$
where $\sigma$ is the electrical conductivity. Thus we can obtain
the electrical resistance $\rho_R =1/\sigma$.  The latter is
directly relevant to the frictional resistance or shear stress
along the tube wall considering the balancing of the driven
electrical force when the transport is finally steady or  fully
developed. This can be evidenced easily once $\chi$ is small then
we have $\rho_R^{-1} \propto\bar{v} \propto \sinh \chi \sim \chi
\propto |-\delta (\rho_e E_z)|$.\newline Fig. 1 shows that there
will be nearly frictionless or almost zero-resistance  states if
we select the activation energy to be $10^{-22}$ J $\sim 0.001$
eV. We can observe a sudden drop of the resistance (frictional or
shear stress) around 3 orders of magnitude at $T=0.12 ^{\circ}$K
($V_h \approx 10^{-24}$m$^3$). It means there is a rather high
electrical conductivity (or very low electrical resistivity)
around this temperatrure for the  material parameters selected.
The qualitative as well as quantative similarity are that this
critical temperature ($T_c$ phenomenon) resembles that found in
amorphous superconductor Mg$_{70}$Zn$_{30}$ (cf.  [9]).
\newline Similarly, Fig. 2 shows that there are almost zero-resistance  states
if we select the activation energy to be $10^{-22}$ J $\sim 0.001$
eV. We can observe a sudden drop of the resistance (frictional or
shear stress) around 3 orders of magnitude at $T=1.9 ^{\circ}$K
($V_h \approx 10^{-23}$m$^3$). It means there is a rather high
electrical conductivity (or very low electrical resistivity)
around this temperatrure for the material parameters selected. The
qualitative as well as quantative similarity are that this
critical temperature ($T_c$ phenomenon) resembles that found in
amorphous superconductor Fe$_{x}$Ni$_{1-x}$Zr$_{2}$ ($x\sim 0.4$;
cf. Table 1 in [10]).
\newline To verify more of our approach, we demonstrate that if we select the activation
energy to be $10^{-22}$ J $\sim 0.001$ eV we can then observe a
sudden drop of the resistance (frictional or shear stress) around
1 order of magnitude at $T=4.56 ^{\circ}$K ($V_h \approx
10^{-23}$m$^3$). It means there is a rather high electrical
conductivity (or very low electrical resistivity) around this
temperatrure for the  material parameters selected. The
qualitative as well as quantative similarity are that this
critical temperature ($T_c$ phenomenon) resembles that found in
amorphous superconductor Zr$_{75}$Rh$_{25}$ (cf. Fig. 1 (a) and
Fig. 2 for the zero (magnetic) field case in [8]).
\newline
Furthermore, Fig. 4 illustrates that there are almost
zero-resistance  states if we select the activation energy to be
$3.5\times 10^{-19}$ J. We can observe a starting (sudden) drop of
the resistance (frictional or shear stress) at $T\sim 164
^{\circ}$K ($V_h \approx 10^{-20}$m$^3$) around 3 orders of
magnitude and below $T=80 ^{\circ}$K. This also demonstrate that
there is a rather high electrical conductivity (or very low
electrical resistivity) around this temperatrure for the material
parameters selected. The qualitative as well as quantative
similarity are that this critical temperature ($T_c$ phenomenon)
resembles that found in amorphous high-$T_c$ superconductor
Hg$_{1-x}$Pb$_x$Ba$_2$Ca$_2$Cu$_3$O$_{8+\delta}$ [Hg($x$Pb)
1:2:2:3] [15].
\newline
To give a brief summary, we already used the verified approach by
borrowing the  analogy from the successful application to the
study of the almost frictionless transport of glassy solid helium
to obtain the critical transport of electric-field-driven
electronic fluids in metallic glasses as well as amorphous (bulk)
superconductors. The critical temperatures we found after  tuning
of the  material parameters resemble qualitatively as well as
quantitatively those superconducting temperatures reported before
in (3D) amorphous  superconductors. Our results show that
effective shear-thinning is an important way to reach
high-temperature charged superfluidity or superconductivity. We
shall investigate other interesting issues [5,16] in the coming
future.

\newpage



\psfig{file=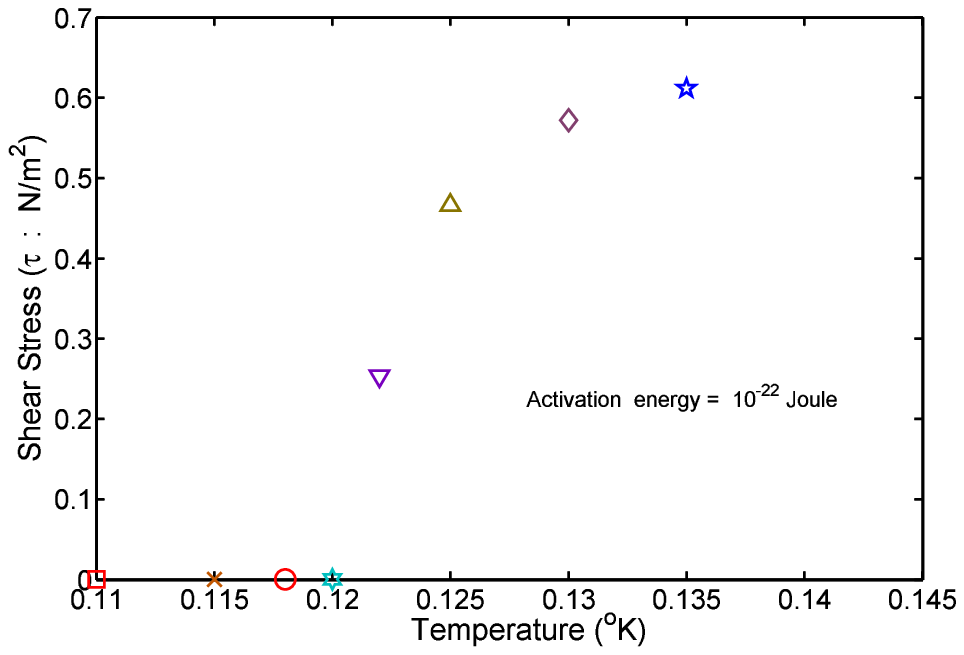,bbllx=-1.5cm,bblly=19cm,bburx=12cm,bbury=27cm,rheight=8cm,rwidth=12cm,clip=}

\begin{figure} [h]
\hspace*{7mm} Fig. 1. Calculated (shear) stresses or resistance
using an activation energy
 $10^{-22}$ J or \newline \hspace*{8mm} $\sim 0.001$ eV. There is a sharp
decrease of shear stress around T $\sim 0.12^{\circ}$K. Around
$0.12$ K, \newline \hspace*{8mm} the electric-field-driven
transport of  (glassy) electronic fluid is nearly frictionless.
\newline \hspace*{8mm} This critical temperature resembles that
found in amorphous superconductor Mg$_{70}$Zn$_{30}$
\newline \hspace*{8mm} (cf. [9]).

\end{figure}

\newpage

\psfig{file=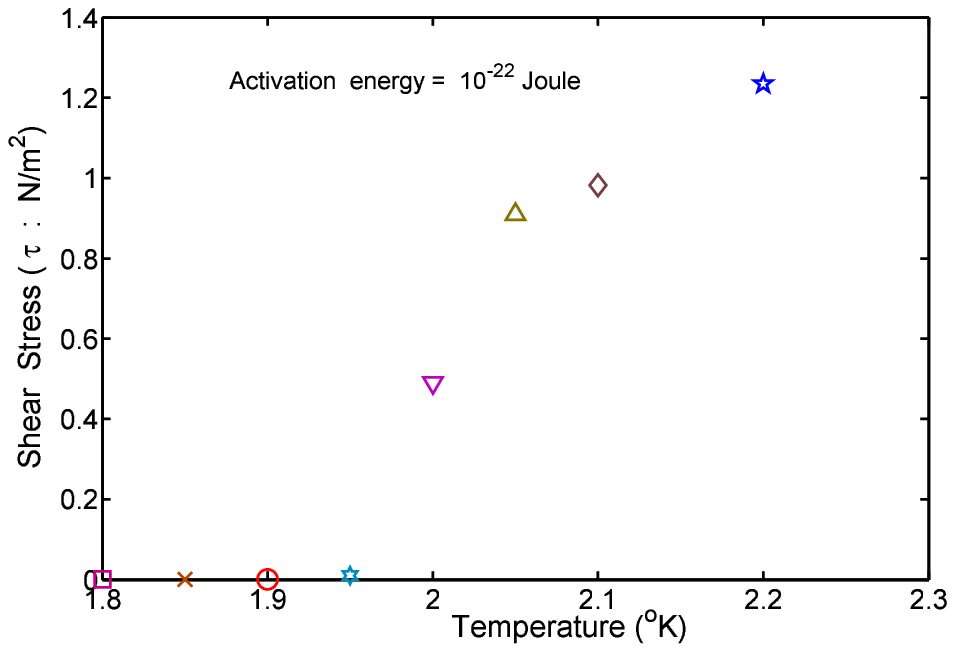,bbllx=-0.5cm,bblly=18cm,bburx=12cm,bbury=27cm,rheight=8cm,rwidth=12cm,clip=}

\begin{figure}[h]
\hspace*{10mm} Fig. 2 \hspace*{1mm}  Calculated (shear) stresses
or resistance using an activation energy
 $10^{-22}$ J or \newline \hspace*{8mm} $\sim 0.001$ eV. There is a sharp
decrease of shear stress around T $\sim 1.9^{\circ}$K. Around
$1.9$ K, \newline \hspace*{8mm} the electric-field-driven
transport of (glassy) electronic fluid is nearly frictionless.
\newline \hspace*{8mm} This critical temperature resembles that
found in amorphous superconductor Fe$_x$Ni$_{1-x}$Zr$_{2}$
\newline \hspace*{8mm} (cf. Table 1 in [10] : $x\sim 0.4$).
\end{figure}

\newpage

\psfig{file=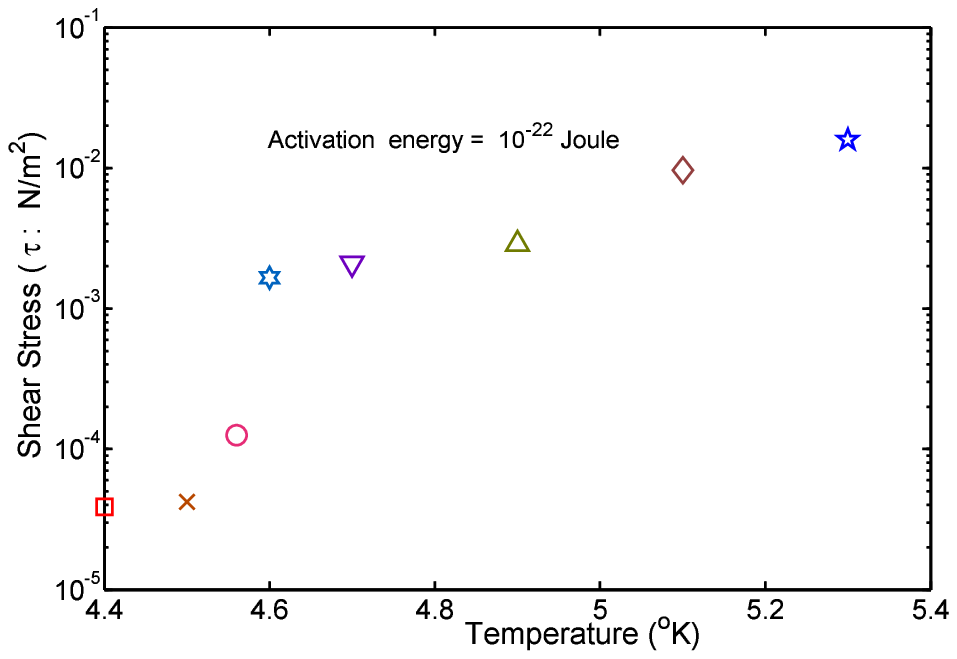,bbllx=-0.5cm,bblly=19cm,bburx=14cm,bbury=27cm,rheight=8cm,rwidth=12cm,clip=}

\begin{figure}[h]
\hspace*{10mm} Fig. 3 \hspace*{1mm} Calculated (shear) stresses or
resistance using an activation energy
 $10^{-22}$ J or \newline \hspace*{8mm} $\sim 0.001$ eV. There is a sharp
decrease of shear stress around T $\sim 4.56^{\circ}$K. Around
$4.5$ K, \newline \hspace*{8mm} the electric-field-driven
transport of (glassy) electronic fluid is nearly frictionless.
\newline \hspace*{8mm} This critical temperature resembles that
found in amorphous superconductor Zr$_{75}$Rh$_{25}$
\newline \hspace*{8mm} (cf. Fig. 1 (a) and Fig. 2 in [8] : the zero (magnetic) field case).
\end{figure}
\newpage

\psfig{file=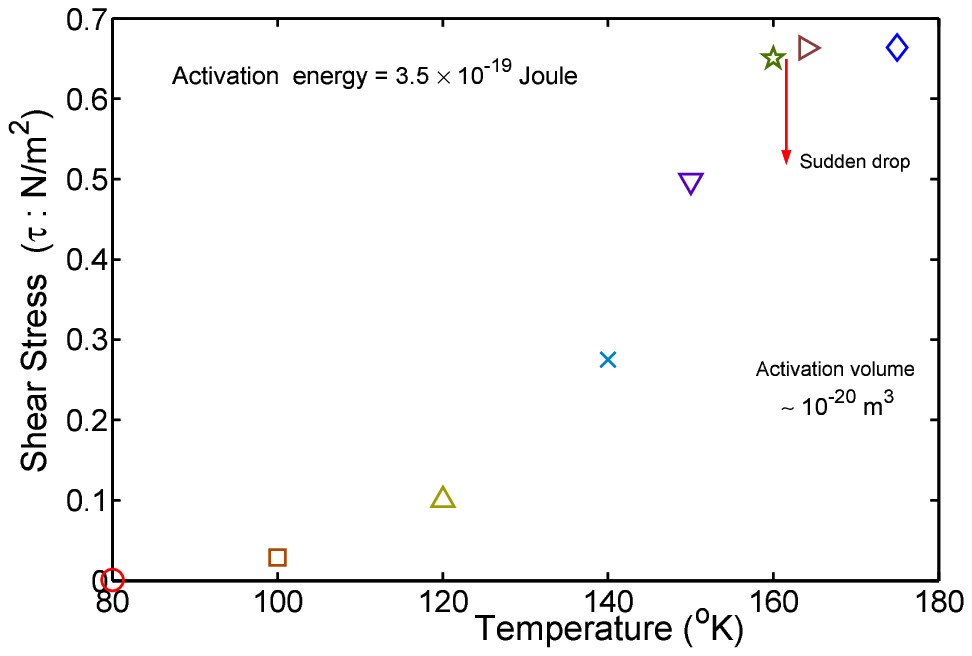,bbllx=0.cm,bblly=19.4cm,bburx=12cm,bbury=28cm,rheight=8.5cm,rwidth=8.5cm,clip=}

\begin{figure}[h]
Fig. 4 \hspace*{1mm} Calculated (shear) stresses or resistance
using an activation energy $3.5 \times 10^{-19}$ J. There is a
sharp decrease of shear stress starting from around $T \sim 164
^{\circ}$K. Below around $80$ K, the electric-field-driven
transport of (glassy) electronic fluid is nearly frictionless.
This critical temperature resembles that found in amorphous
high-temperature superconductor
Hg$_{1-x}$Pb$_x$Ba$_2$Ca$_2$Cu$_3$O$_{8+\delta}$ [Hg($x$Pb)
1:2:2:3] [15].
%

\end{figure}
\end{document}